\documentclass[titlepage,12pt]{article}\usepackage[]{graphicx}\usepackage[]{color}
\makeatletter
\def\maxwidth{ %
  \ifdim\Gin@nat@width>\linewidth
    \linewidth
  \else
    \Gin@nat@width
  \fi
}
\makeatother

\definecolor{fgcolor}{rgb}{0.345, 0.345, 0.345}
\makeatletter
\@ifundefined{AddToHook}{}{\AddToHook{package/xcolor/after}{\definecolor{fgcolor}{rgb}{0.345, 0.345, 0.345}}}
\makeatother

\usepackage{framed}
\makeatletter
 {\par\unskip\endMakeFramed%
 \at@end@of@kframe}
\makeatother

\definecolor{shadecolor}{rgb}{.97, .97, .97}
\definecolor{messagecolor}{rgb}{0, 0, 0}
\definecolor{warningcolor}{rgb}{1, 0, 1}
\definecolor{errorcolor}{rgb}{1, 0, 0}
\makeatletter
\@ifundefined{AddToHook}{}{\AddToHook{package/xcolor/after}{
\definecolor{shadecolor}{rgb}{.97, .97, .97}
\definecolor{messagecolor}{rgb}{0, 0, 0}
\definecolor{warningcolor}{rgb}{1, 0, 1}
\definecolor{errorcolor}{rgb}{1, 0, 0}
}}
\makeatother

\usepackage{alltt}

\usepackage{natbib}
\usepackage{caption}
\usepackage[latin1]{inputenc}
\usepackage{graphicx}
\usepackage{colortbl}
\usepackage[space]{grffile}  
\usepackage{authblk}
\usepackage{url} 
\usepackage{amsmath}
\usepackage{hyperref}  
\def\spacingset#1{\renewcommand{\baselinestretch}%
{#1}\small\normalsize} \spacingset{1}

\newcommand{\equationN}{\addtocounter{eqnum}{1}\begin
{equation}}
\newcommand{\eqnarrayN}{\addtocounter{eqnum}{1}
\begin{eqnarray}}

\newcommand{\E}{\mbox{\rm E}}
\newcommand{\Prp}[1]{\mbox{\rm P\hskip-0.25em}\left(#1\right)}

\newcommand{\var}{\mbox{\rm var}}

\newcommand{\corr}{\mbox{\rm corr}}

\newcommand\indep{\perp \!\!\! \perp}

\newcommand{\pihat}{\hat{\pi}}
\newcommand{\muvec}{\mbox{\boldmath $\mu{}$}}
\newcommand{\evec}{\mbox{\boldmath $\eta{}$}}
\newcommand{\thvec}{\mbox{\boldmath $\theta{}$}}
\newcommand{\xvec}{\mbox{\boldmath $X{}$}}
\newcommand{\Yvec}{\mbox{\boldmath $Y{}$}}
\newcommand{\ytvec}{\Yvec^T}
\newcommand{\yovec}{\Yvec^{(0)}}
\newcommand{\yonevec}{\Yvec^{(1)}}
\newcommand{\yltvec}{\Yvec^{(t)}}
\newcommand{\ybltvec}{\Yvec^{(0:t)}}
\newcommand{\svec}{\mbox{\boldmath $S{}$}}
\newcommand{\svecs}{\mbox{\boldmath $S^*{}$}}
\newcommand{\tvec}{\mbox{\boldmath $T{}$}}

\IfFileExists{upquote.sty}{\usepackage{upquote}}{}
\begin{document}

\spacingset{1.45} 
\title{\bf A note on  stratification errors in the analysis of clinical trials}
\author[1]{Neal Thomas}
\date{{July 2023}}
\maketitle

\begin{center}
{\bf Abstract}
\end{center}

Stratification in both the design and analysis of randomized 
clinical trials is common.
Despite features in automated randomization systems to re-confirm the 
stratifying variables, incorrect values of these variables may be entered.   These
errors are often detected during subsequent data collection and verification.
Questions remain about whether to use the mis-reported initial stratification
or the corrected values in subsequent analyses.   
It is shown that the likelihood function resulting from the design of  randomized clinical trials
supports the use of the corrected values.  New definitions are proposed that characterize
misclassification errors as `ignorable' and `non-ignorable'.
Ignorable errors may depend on the correct strata and any other
modeled baseline covariates, but they are otherwise unrelated to potential treatment outcomes.
Data management review suggests most misclassification errors are 
arbitrarily produced by
distracted investigators, so they are ignorable or at most weakly dependent on
measured and unmeasured baseline covariates.  
Ignorable misclassification errors may produce a small increase in standard errors, but 
other properties of the planned analyses are unchanged  
(e.g., unbiasedness, confidence interval coverage).
It is shown that unbiased linear estimation in the absence of misclassification errors 
remains unbiased when there are non-ignorable  misclassification errors,
and the corresponding confidence intervals based on the corrected strata values are conservative.

\vskip1in\noindent{\bf Acknowledgment}  The author thanks Roy Choudhury 
for his generous suggestions.

\pagebreak
\section{Introduction}
\label{intro}
Clinical trials are routinely stratified in their design with 
randomization based on the initially reported strata, which may contain errors that
are subsequently corrected before database closure. We assume that 
the randomization does not depend on other patient characteristics,
even if some covariates are planned for the analysis in addition to stratification
indicators.  
The randomization method considered is permuted blocks nested within the initially entered strata.   For example, it is common in trials comparing two treatments, 
`A' and `B', to assign patients to treatments based on the order of random permutations of $(A,A,B,B)$.  Each `block' of size $4$ is nested within a stratum
to achieve an equal number of the two treatments within each stratum.  
The number in each treatment group may not be exactly equal if 
there are partially filled blocks when the trial ends.  
When the investigator performing the randomization enters an incorrect value for a stratum, the 
patient is assigned to the next available treatment code in the incorrect
stratum rather than the next available treatment code in the correct 
stratum.   Our experience has been that the
rate of misclassifications is typically $1-2$\%, but rates as high as $10$\%
have been observed in a large trial with complex stratification factors.  
Higher rates of
misclassifications might occur if the patients are stratified based on
 rapidly evaluated laboratory measurements
that are more precisely updated later \citep{ke2017}.  
Successful blinding that prevents investigators and sponsors from
predicting future treatment codes in the randomization lists is assumed until the concluding section.  Likelihood methods
are considered because they are especially useful for creating estimators and test
statistics.  
Randomization-based inference in which all probabilities are derived from
the randomization design is also considered 
\citep{ludbrook98,rosenberger02,ke2017}.

A minimal modeling framework is developed with only the 
essential features necessary to identify 
conditions for valid inference.  Endpoint types within this framework
include continuous, binary and time-to-event.   
Stratification refers to an exhaustive non-overlapping
classification based on baseline characteristics, which may be formed by crossing
several factors such as gender and baseline disease severity. 
Stratification in design means
the randomization is implemented utilizing separate 
independent randomization lists 
within each stratum for the
purpose of achieving pre-specified proportions of each treatment group within each
stratum.  Stratification in the analysis, which is also pre-specified, refers to the 
use of covariate-adjusted models that condition on strata
indicators when comparing treatments.  Stratification in the analysis may also be 
implemented by forming strata-specific treatment effect estimates that are 
then combined into an overall treatment effect 
estimate using weights.  Analysis-stratification may be
applied with a stratified or unstratified design-randomization.  
Different sample size per 
treatment group (e.g., 2-1 versus 1-1)
can be targeted, but it is assumed here that the targeted treatment allocation does not
vary across strata. This restriction is met in most clinical trials. Provided this
restriction is maintained, our results apply with differing block sizes 
randomly selected, but the discussion and examples will focus on the common
fixed block size design.  
Compliance with the treatment assigned is assumed, even if a different treatment
might have been assigned in the absence of stratification errors.  
The analysis must be fully pre-specified.

Misclassification will be characterized as ignorable or non-ignorable.   
This characterization is closely analogous to the usage of these
terms in the missing data \citep{littlerubin19} and causal inference
 \citep{rubinimbens15} literature.  
Ignorable misclassification, which is covered in Section \ref{ignsect}, occurs
when classification errors may depend on the correct strata and any other
modeled baseline covariates, but is otherwise unrelated to potential treatment outcomes.
A very general form of the likelihood is considered, which is developed
from notation introduced in Section \ref{notation}, so the recommendation
to use the corrected strata in the likelihood function with ignorable misclassification
applies regardless of endpoint type, estimation of interactions, estimation of
conditional or unconditional parameters, estimation of non-linear model parameters,
and with or without the inclusion of a prior distribution.  The implications of non-ignorable
misclassification are considered in Section \ref{nonignsect}.  Non-ignorable 
misclassification occurs when the presence of classification errors is predictive of 
potential outcomes even after accounting for the correctly recorded baseline variables 
included in the model.  
	Surprisingly, non-ignorable errors do not create bias in otherwise unbiased 
linear estimators, and the 
confidence intervals associated with these estimators become conservative.   This result
is demonstrated with some simulations of extreme non-ignorable settings in Section \ref{simnon}.  
Existing theory is then used to explain these findings in 
Section \ref{theorysect}.

\section{Likelihood function with ignorable misclassification}
\label{ignsect}

\subsection{Notation}
\label{notation}

Notation for a stratified trial with $N$
patients, a control, and $t$ treatments begins with 
$t+1$ potential outcomes for  
the control and $t$ treatments for the
$i^{th}$ patient, denoted by $y_i^{(0)},y_i^{(1)},\hdots,y_i^{(t)}$.  
This endpoint notation
includes continuous, binary, and time-to-event endpoints (including
censoring variables if needed).  It could be multivariate, though multiplicity
issues are not considered.  The treatment assigned and 
received is denoted by
$T_i=0,1,\hdots,t$.  The observed response for the $i^{th}$
patient is given by $y^{T_i}_i$.  
The notation and description of the potential outcomes follows that of the 
causal inference literature \citep{rubinimbens15,hernanrobins23}.  
The stratification variable is denoted by
$S_i=1,\hdots,m$, where $m$ is the number of exhaustive non-overlapping
stratifying subgroups.  Stratification values entered into the randomization system by
the investigators at clinical sites, which may include errors, are denoted by $S^*_i$.   
Any additional baseline covariates planned for the
primary analysis are denoted by $x_i$.  The variables representing the values from individual
patients are collected into vectors and matrices representing the entire samples:
$\ytvec$, $\svec$, $\xvec$, $\svecs$,  and the potential outcomes 
$\ybltvec=(\yovec,\yonevec,\hdots,\yltvec)$.

\subsection{Likelihood function}

\label{likf}
The likelihood function will be constructed by factoring a sequence of conditional 
distributions including an additional
conditional distribution for the mistaken values, $\svecs$.  The distribution
of the outcomes conditional on the treatment assignments, stratifiers, and other covariates
is typically of primary interest, with parameter(s) denoted by $\muvec$ (e.g., difference
in treatment means, log-odds, hazard ratios).  The distribution of the stratifying variable
and other potential covariates is most often viewed as ancillary \citep{cox74} and 
it is not estimated, but there are exceptions such as the reporting of the risk difference, which
 may require either the implicit or explicit estimation of the distribution of the baseline
stratifier and other covariates.  The parameters of the distribution of $\svec$ and $\xvec$ are denoted
by $\thvec$.  The $\thvec$ can be estimated assuming that the
trial participants represent a random sample of some population, or using external
epidemiology data for the baseline covariate distributions \citep{heitjan21,wang19}.  The distribution
of $\svecs$ conditional on $\svec$ and $\xvec$ may depend on unknown parameters such as the
rate of random misclassifications, denoted
by $\evec$.  The full parameter space is the cross-product of the appropriate domains for
the parameters $\muvec$, $\thvec$, and $\evec$.

The randomization ensures the distribution of $\Prp{\ytvec\mid \tvec, \svec, \xvec, \svecs}$
is formed from the intended causal distributions 
$\Prp{\yovec\mid  \svec, \xvec, \svecs}$,   
$\Prp{\yonevec\mid  \svec, \xvec, \svecs},\ \hdots$\ ,\hfil\break
$\Prp{\yltvec\mid \svec, \xvec, \svecs}$, 
for patients assigned to control and the different treatments, 
which is assumed here without further discussion
\citep{rubinimbens15,hernanrobins23}.   The likelihood function is given by

\begin{eqnarray}
\Prp{\ytvec, \tvec, \svec, \xvec, \svecs \mid \muvec, \thvec, \evec} 
&=& \Prp{ \ytvec \mid \tvec, \svec, \xvec, \svecs, \muvec, \thvec, \evec}
\Prp{ \tvec \mid \svec,  \xvec, \svecs,\muvec, \thvec, \evec}\times \nonumber\\
&&\Prp{ \svecs \mid \svec, \xvec, \muvec, \thvec, \evec }
\Prp{ \svec, \xvec \mid  \muvec, \thvec, \evec }
\end{eqnarray}
Based on the parameter definitions and the fact that treatment assignment depends
only on $\svecs$, the likelihood simplifies to

\begin{eqnarray}
\Prp{\ytvec, \tvec, \svec, \xvec, \svecs, \mid \muvec, \thvec, \evec} 
&=& \Prp{ \ytvec \mid \tvec, \svec, \xvec, \svecs, \muvec}
\Prp{ \tvec \mid \svecs}\times \nonumber\\
&&\Prp{ \svecs \mid \svec, \xvec, \evec }
\Prp{ \svec, \xvec \mid  \thvec }\nonumber\\
&\propto& \Prp{ \ytvec \mid \tvec, \svec, \xvec, \svecs,  \muvec}
\times \nonumber\\
&&\Prp{ \svecs \mid \svec, \xvec, \evec }
\Prp{ \svec, \xvec \mid  \thvec }  \ .
\label{lik2}
\end{eqnarray}
The parameters, $\evec$, are not of primary interest.  Because $\evec$ only appears in one
multiplicative term in the likelihood
(\ref{lik2}), the term with $\evec$ can be discarded yielding
\begin{equation}
\Prp{\ytvec, \tvec, \svec, \xvec, \svecs \mid \muvec, \thvec}
\propto \Prp{ \ytvec \mid \tvec, \svec, \xvec, \svecs,  \muvec}
\Prp{ \svec, \xvec \mid  \thvec } \ .  \label{lik3}
\end{equation}

\subsection{Ignorable misclassification}

Discussion with clinical investigators and study monitors indicate that 
most misclassification errors are arbitrary recording mistakes made by 
distracted investigators that do not depend on patient characteristics, so a 
`random mistake' assumption is reasonable.  For the situation where
less precise preliminary
measurements are used to compute strata values, 
the noisy measurement does not improve prediction of the potential outcomes 
once the
more precise measurements are known.  Stated as statistical independence,  
the mistakes may depend on the correct values of the stratification (and any other
baseline covariates included in the analysis model), but they do not depend
on other patient characteristics related to the potential outcomes:
\begin{equation}
\ybltvec \indep \svecs \mid \svec, \xvec \ , 
\label{ignore}
\end{equation}
where $\indep$ means independent. 
This conditional independence will be called ignorable misclassification, because
the resulting likelihood function for $(\muvec,\thvec)$ then depends only on 
the corrected strata values and not the incorrect values.  
Intuitively, when $\svec$ and $\xvec$ are known, 
$\svecs$ determines treatment assignments based on 
noise similar to a random
number generator, so they do not confound treatments, $\tvec$, with any other covariates, 
measured or unmeasured, and they do not improve prediction of $\ytvec$, so they can
be discarded.  Appendix \ref{appA} contains a formal derivation showing that
the likelihood takes the form
\begin{equation}
\Prp{\ytvec, \tvec, \svec, \xvec, \svecs  \mid \muvec, \thvec} 
\propto \Prp{ \ytvec \mid \tvec, \svec, \xvec, \muvec}
\Prp{ \svec, \xvec \mid  \thvec }\ .
\label{uform}
\end{equation}
Several authors have evaluated the consequences of misclassified strata.  
\citep{ke2017,fan18} and \citep{yelland23} primarily used simulations to evaluate
misclassification on different response types, and \citep{wang21} used 
asymptotic calculations evaluating linear models for continuous outcomes.
All of the simulations and theory in these references 
have ignorable misclassifications. This is analogous to
missing at random (MAR \cite{littlerubin19}), 
because the misclassifications
depend at most on the patients' strata through 
different rates of shifting from/to the correct strata.  Most of the
simulations were analogous to missing completely at random (MCAR) because
the misclassification error rates were the same regardless of the 
correct strata.  
The Type I error rates,
power, and other measures of inferential performance for
the methods applied with the correct strata in
their simulations were excellent (e.g., Tables 5 and 6 of \citep{ke2017}).  

The performance they report  
with ignorable misclassification is due to the optimal large sample properties of 
likelihood-based inference.  Their estimators and standard errors evaluated are 
maximum likelihood methods, and conditional on the treatment
assignments and correct strata, the results were all generated from
independent identically distributed covariates, outcomes, and misclassification
errors.  This remains the case even though the permuted block assignments 
create dependence between treatment indicators, and between treatment and 
strata indicators.  The likelihood result is very general and also applies to response
types not evaluated in the references.  The ignorable misclassification of strata
reduces the efficiency of the stratified design, but it does not change the 
optimal analysis.  \cite{fan18} and \cite{yelland23} included simulations that
evaluate sensitivity of the various analyses to misclassification errors that
may remain after the known errors are corrected.  Sensitivity analyses
like these are not covered here.  The derivations in \cite{wang21} assume none
of the misclassification errors are detected and corrected, so they focus on
unconditional analyses and analyses that condition on the strata with 
mis-classifications,  thus their results are not directly applicable in the 
setting considered here.

It is important to distinguish randomization-based procedures that utilize the
incorrect strata, $\svecs$, in both the test statistic and in the randomization
probabilities, from those that use the correct strata,
$\svec$, in the test statistic but use $\svecs$ 
when computing the randomization
distribution. The latter often involves simulation-based assessment of the distribution.  
Confidence intervals and estimators derived by inverting test statistics
that replace $\svec$ with $\svecs$ may not correctly cover strata-specific parameters. 
An important example of this failure is estimation of
the difference in treatment means within a strata (subgroups, \citep{ke2017,yelland23}). 
Inference based
on test statistics using the incorrect values differ from both
the randomization-based and corresponding 
likelihood-based procedures that use the correct values except in settings where
the target summary of a distribution is the same 
with and without conditioning on covariates.

\section{Ignorable and non-ignorable misclassification}
\label{nonignsect}

While the assumption of ignorable misclassification is reasonable, it cannot be assured 
because the misclassification errors are not controlled.
Even when blinding is effective and the errors are arbitrary, incidental dependence with
the potential clinical outcomes is possible.  For example, if a few clinical investigators
are prone to these errors, and they are at sites that 
recruit healthier patients,   
$\svecs$ could be predictive of positive outcomes not predicted by $\svec$.
Without the ignorable misclassification condition, the key term in the likelihood 
retains the form, $\Prp{ \ytvec \mid \tvec, \svec, \xvec, \svecs, \muvec}$.
Broadly based guidance is to include $\svec$ and $\svecs$ as covariates in the adjustment
model because treatment assignement depends on $\svecs$, and the
$\svec$ are predictive of the potential outcomes 
(strongly ignorable treatment assignment, \citep{rr83,shao2010,
bugni18,ye2020}).  
In practice, misclassification rates of approximately 
one percent are typical, 
so the inclusion of
both $\svec$ and $\svecs$ in the model is 
not feasible.  Inclusion of both, or $\svecs$ alone,  
would also yield a conditional model that
is difficult to interpret and use. 

Simulations are used in the next three sections 
to demonstrate that estimation stratified on the corrected values yields unbiased estimation 
and conservative coverage of confidence intervals in some extreme settings where 
incorrect biased estimation may seem inevitable. 
The simulation results are then
explained in Section \ref{theorysect} using existing theory applied in the non-standard  setting of misclassification.

\subsection{Simulation design}
\label{simsect}

An eighty-patient study with a $1:2:2$ allocation to control and two
treated groups was simulated with two strata.  An
unequal treatment allocation and a
third group were added because their presence can adversely impact the 
properties of some model-based estimation.  
Strata membership (low/high baseline severity) for each 
patient is determined independently 
with probabilities $0.4$ and $0.6$, so the number of patients in the strata vary across
simulation replications.  Normally-distributed outcomes were generated with a substantial
strata difference:  $y^{(0)}\sim N(0,1)$ and $y^{(0)}\sim N(1,1)$ in the 
low and high strata, respectively.
An additive treatment effect is simulated for both 
active treatments:  $y^{(1)}\sim N(0.5,1)$ and 
$y^{(1)}\sim N(0.5+1,1)$, in the low and high strata, respectively, and likewise
for the potential outcomes to the second active treatment, 
$y^{(2)}\sim N(0.5,1)$ and 
$y^{(2)}\sim N(0.5+1,1)$ in the two strata.  
Although only one value will be observed, a triplet of 
$(y^{(0)},y^{(1)},y^{(2)})$ are simulated for each
patient.  The initial simulations, consistent with  
randomization-based inference, 
set the correlations between each pair of the potential outcomes equal to $1.0$.  
Additional simulations create 
heterogeneous treatment effects by setting the 
correlations between each pair of the potential outcomes to $0.5$.  
The treatment assignments, $\tvec$, are assigned in permuted blocks of size 
$10$.  There is no
heterogeneity or dependence in the potential outcomes between the blocks  
within the strata.  In real trials, heterogeneity 
in patient recruitment may produce heterogeneity between randomization blocks
over time that results in conservative 
confidence intervals by creating positive within-block 
correlation \citep{demets15,gansky94,matts88}.  The nesting of blocks
within calendar time is a ubiquitous feature of permuted block designs.  
The resulting conservative
confidence intervals can be viewed as a situation where 
stratification in design is ignored in analysis.  The derivation 
of the likelihood function
in Section \ref{ignsect} similarly ignored calendar time, which is 
easily justified only when calendar
time is assumed to be independent of the potential outcomes.

Treatment effects were estimated using a simple linear model 
that includes 
separate additive effects across the strata for each 
treatment group, an intercept and
an indicator variable for the strata.  
A homogeneous variance is fit across each combination of 
the treatments and the strata.  The treatment effects 
are estimated by least squares 
with confidence intervals and null tests formed from the $t-$distribution.  
These assumptions are consistent with the 
initially simulated data, and they were used throughout 
so that any distortion
in these assumptions due to the misclassfication in 
the randomization strata 
would diminish their performance.  Results are presented
using both the correct strata and the misclassified strata 
in the estimators and test statistics.  
The randomization distribution was always computed 
using the observed (possibly misclassified) strata.
  
\subsection{Simulation of ignorable misclassification}
\label{simig}
 
Simulations ($400,000$ replications) with ignorable  
misclassifications are summarized in the upper section of
Table \ref{simtab}, which includes estimates
of bias and coverage of nominal $95$\% confidence intervals
for the treatment effect estimation.  
The standard deviation of the simulation estimates of the 
coverage probabilities in Table \ref{simtab} 
is approximately $0.0007$. 
The table summarizes the 
estimation of the mean difference between the
first treatment and control.  By symmetry, the same 
results apply to the second treatment.   
Simulations with random misclassification rates of $\gamma_L=\gamma_H=2$\% in each strata were
performed, and then the misclassification rates were increased to $\gamma_L=15$\% in
the lower strata and $\gamma_H=30$\% in the upper strata.  
The estimator stratifying on the corrected values is 
unbiased with the intended confidence coverage in all of the
simulations with ignorable misclassification.  The upper
section of Table \ref{simpowtab} summarizes the power and level
of the null hypothesis test under the same ignorable conditions.  
The simulation standard errors are $\leq 0.008$. These errors
are larger than in Table \ref{simtab} 
as they are based on $4000$ simulation replications 
because randomization-based testing was added, which required
simulation of the randomization distribution within each
simulation replication.  
The level of the test matches the $0.05$
level within simulation error when using likelihood or
randomization based inference stratifying on the correct 
strata in the test statistic.  Results based on the linear
model estimator formed from the misclassified strata are
also reported in Table \ref{simpowtab}.  The likelihood
and randomization based inference were in good agreement 
when using correct or misclassied strata in the estimator.  

The population mean and standard deviation ($SD$) by treatment group within the 
ignorably misclassified
strata are given in the upper portion of Table \ref{poptab}.  
The ignorable misclassifications resulted 
in more variability within the incorrect randomization 
strata, so   
analyses using the misclassified strata were less precise,
as displayed by the mean $SE$ in Table \ref{simtab} and
power in Table \ref{simpowtab}.  With low misclassification rates
(e.g., $\leq 2$\%), the impact is low but it can be appreciable
if the misclassification rate is higher.  
In the homogeneous effect setting for example, the mean 
standard errors ($SE$) for 
the estimator stratifying on the 
incorrect strata values are $0.31$ and
$0.332$ for the low and high misclassification simulations, respectively,
while the corresponding mean $SE$ when stratifying on the corrected values are
$0.307$ and
$0.308$.  
The mean $SE$ for 
stratified estimation based on 
an unstratified permuted block design is also $0.307$ in 
the simulated example (not displayed).
If it is anticipated that an
analysis with the incorrect strata will be required, perhaps as a `sensitivity' analysis that 
often means the least favorable result will be utilized, 
pre-specifying an analysis adjusting for the strata without the design stratification
should be considered for larger studies due to the low benefit of design
stratification.  

\subsection{Simulations with non-ignorable misclassification}
\label{simnon}

Extreme non-ignorable misclassification was
simulated by misclassifying all patients in the low strata with 
$y^{(0)}\geq F_0^{-1}(1-\gamma_L)$, where $F_0$ is the cumulative distribution function (CDF) of
$y^{(0)}$ in the low strata, and
misclassifying all patients in the high strata with 
$y^{(1)}\leq F_1^{-1}(\gamma_H)$ where $F_1$ is the CDF of
$y^{(1)}$ in the high strata.  
Investigators can not observe the potential outcomes when assigning
strata, but this extreme situation could be approximated by investigators 
intentionally misclassifying strata
using highly predictive baseline covariates not available to the data analyst. Note that these
misclassifications depend on the correct strata and the potential 
outcomes, but not the potential treatment
assignments, $T$. 

The treatment-specific 
population means and SD's within the non-ignorably misclassified strata 
when there is high misclassification are in the
middle portion of Table \ref{poptab}.  Between  
the two misclassified strata,  the treatment effect
is no longer additive when the correlation between the
potential outcomes is $0.5$, although the overall mean 
treatment difference remains unchanged.
These non-ignorable misclassifications were constructed so that the strata used in the randomization had
less variability than the correct strata.  An additional set of misclassified strata
were simulated by reversing the misclassifications so that patients in the low strata
were misclassified when $y^{(0)}\leq F_0^{-1}(\gamma_L)$ and patients in the high strata
were misclassified when $y^{(1)}\geq F_1^{-1}(1-\gamma_H)$. This misclassification produced
many resulting misclassified strata with more 
variability than the original correct strata. These are
summarized in the lower portion of Table \ref{poptab}.

The likelihood-based stratified estimation using the 
correct strata remain unbiased for both non-ignorable misclassification 
settings, and 
the coverage of the confidence intervals is conservative (Table \ref{simtab}).  The amount of over-coverage varies 
and its relationship to the misclassification is complex.  The
corresponding level of null hypothesis testing is also 
conservative along 
(Table \ref{simpowtab}).  When using the correct strata in the analyses,
the power was generally higher than for the corresponding ignorable
setting because the misclassified strata were predictive and imbalance
in these strata were reduced by the design.  The randomization-based
test accounts for this improved balance, but it is not included in
the likelihood-based analyses yielding lower power for them.
With randomization-based tests, the level of 
the testing using the correct strata is 
consistent with the $0.05$ value, and the power of the tests
are improved.  When the misclassified strata are used in the 
estimator and test statistic, the coverage of confidence 
intervals and the levels of the test are also consistent with the
intended values.  The performance of the likelihood and randomization based methods was indistinguishable 
when applied to the misclassified strata (Table \ref{simpowtab}), despite
violations of the assumptions of a common treatment
effect across misclassified strata and homogenuous variance
across strata and treatment groups (Table \ref{poptab}). 

Comparison of the power of the tests using the correct versus the misclassified
strata in the test statistic can be appreciably improved or
degraded depending on the variability in the 
misclassified strata, demonstrating there is no simple 
global claim for superiority of either option at the
time of study design.  Reasons for preferring the strata
based on the correct values are 1) face validity with
non-statistical users of the study results, 2) consistency
in the use of the correct values between overall treatment
effect estimation, subgroup estimation, and 
estimation of other conditional estimands, and 3)  the
correct strata are superior when the misclassfication is 
ignorable, and approximate ignorability is the most plausible
condition.

\subsection{Theory applicable to non-ignorable misclassification}
\label{theorysect}

Justification from likelihood theory to `analyze as randomized' suggests 
it may be sufficient to use $\svecs$ alone due to the independence
of $\svec$ and treatment assignment, $\tvec$, conditional on $\svecs$.  
The guidance is problematic
in this setting because if $\svec$ is not included, it is 
not known which values in $\svecs$ are incorrect.  Appropriate
modeling of $\ytvec$ given $\svecs$ and $\tvec$ thus requires latent mixture models, can
be inefficient especially when the misclassifications create outliers, and it may target
unintended estimands (e.g., invalid subgroup estimation).  It also produces a qualitatively
different estimation approach from the setting with ignorable misclassification that is the
most plausible approximate condition 
when a few sporadic operational errors are present.  
Alternatively, to justify the use of $\svec$ only, 
a more detailed assessment is required that 
utilizes features of the 
stratified randomization design.  

The propensity score is a method to identify and construct a small subset of covariates from 
a larger
collection of baseline covariates that have 
a `balancing' property permitting valid inference
to be constructed from the small subset of covariates 
whenever valid inference could be achieved using the larger collection (\cite{rr83}).  The
propensity scores, which are the $\Prp{T_i \mid \svec, \xvec, \svecs}$, are
constant for the permuted block implementations of
stratification. To simplify the discussion, it is assumed here that the treatment
effect estimation and resulting propensity score are restricted to comparison of
one of the treated groups and control, although multivariate versions of the
propensity score can be defined.  
The propensity scores do not
recover the stratification, $\svecs$, because the construction of
propensity scores includes an under-appreciated assumption that the individual $T_i$ in $\tvec$
are independent given the covariates, which is not the case with 
permuted block designs
(Section 1.2, \cite{rr83}).  While the propensity scores do not yield a 
fully satisfactory summary,
they reveal the important fact that the $\Prp{T_i \mid \svec, \xvec, \svecs}$ 
is constant, i.e., each $T_i$ is independent of all baseline 
values.  As a consequence,
linear estimators such as the
difference in means or proportions, both stratified on the
correct strata or unstratified, are unbiased when they 
would be unbiased in the absence of 
misclassificaton errors because their expectations 
depend on the expectations of the $T_i$, but not their 
joint distribution.  This result was developed using direct
calculation by \citep{ke2017}
assuming ignorable misclassification 
and the simple linear model used in the 
simulation Sections \ref{simsect}-\ref{simnon}.  
They also note similar results were obtained
by \citep{gust02} using a measurement 
error approach.  The propensity
score approach highlights the fact that the covariate distribution
is the same for patients assigned to each treatment 
as the reason that bias is not introduced by the misclassification.

There is research evaluating the impact of ignoring design stratification
when computing standard errors and confidence intervals for 
treatment effects (e.g., \citep{shao2010,bugni18,ye2020,fda23}).
Ignoring design stratification in the analysis 
yields conservative inference, i.e., the 
estimated standard error 
is at least as large as the actual standard error,  and
the coverage is greater than or equal to the nominal
interval coverage.  This is a
broadly recognized finding, but its application to
stratification based on $\svecs$ especially when $\svecs$ contains information
about response beyond that in $(\svec,\xvec)$ may be surprising.  
\citet{bugni18} showed the stratified estimator 
is also conservative,
but  this result is not directly applicable when there is 
misclassification.

The results in \citet{bugni18} can be extended to estimation of 
the treatment effect within a pre-specified subgroup.    
Details of the derivation are in Appendix \ref{appb}.
Subgroup analyses
ignoring stratification are common, so the fact
that these analyses are conservative is an important 
corollary of the results in \cite{bugni18}.  
The subgroup result also clarifies the consequence of using 
a different stratification in the design than in the analysis.  
When the subgroup is an  
analysis stratum, it shows
the variance calculation for the treatment effect estimator 
within each analysis stratum
ignoring the design stratification is
conservative.  Linear stratified estimators 
are  weighted averages of the strata 
specific estimators, thus a stratified analysis using strata that differ
from the design strata will yield conservative inference.  The application
of most interest here utilizes the corrected strata in the analyses but
the (potentially) informative misclassified strata in the design.  
This heuristic is not a rigorous
proof because it does not account for potential dependence
between the strata-specific estimators.  A more detailed
derivation is not developed here.   

The literature referenced here 
does not cover the general case of non-linear parameter estimation.    
The results in \cite{ye2020} establish that the unstratified log
rank test is convervative when applied to a stratified design, which is
a common and important example of the more general approach.  
Simulation studies (e.g., \cite{fan18}) suggest similar results with other models non-linear 
in their parameters.  As with
the conservative results produced by ignoring permuted 
blocks in the analysis \citep{matts88}, 
the results in \citep{shao2010,bugni18, ye2020} require 
that patients within  strata are more similar than between strata.  
This is implicitly implied in these references 
and in the simulations reported 
in Sections \ref{simsect}-\ref{simnon} by assuming the 
patients are independent, and identically distributed 
within strata.  As with the  unstratified permuted 
block designs, this assumption cannot be guaranteed, but it would
be unexpected to find less variability in the 
differences between 
patients selected within strata  
than differences between patients selected randomly from 
the full study sample.

\section{Conclusions}
\label{conclusions}
The impact of misclassifying a small percent of the 
strata values that are
subsequently corrected when these errors are unrelated to patient characteristics or have at most a weak incidental dependence
is inconsequential except in small trials where there may be some loss in efficiency due to 
diminished design stratification. As noted in Section 3.3, 
mathematical examples of non-ignorable misclassification 
conditions can be created that favor either the use of the corrected 
or misclassified strata in the analysis, but the corrected
values are recommended because they are superior under
the most plausible condition of ignorability, and
they are conservative in the settings in which they are 
not optimal.  A more complete characterization of the impact of 
stratification and randomization approaches in the setting of non-ignorable 
misclassification of strata, especially in the
setting of non-linear parameter estimation would 
further strengthen recommendations regarding non-ignorable misclassification.  
It is nearly ubiquitous in current clinical trial analyses
to ignore the blocking
on time of entry into clinical trials entailed by the use of permuted block 
randomization designs.
The issue of design stratification that differs from the analysis stratification 
is not limited to the misclassification of strata. 

More concerning for validity of results are attempts by 
investigators or the sponsor to manipulate the
assignment of treatments when the study is not  
effectively blinded,  The lack of blinding may be due
to the infeasibility of concealing treatments
once the treatment has begun, or more severely
when someone has access to the randomization
lists and acts with fraudulent intent.
The potential for manipulation in the first situation, intentional or not, 
has been evaluated by many authors
following \cite{blackwellhodges57}, who recommend large block sizes to
reduce predictability in studies without effective blinding.  
If future treatment assignments in each strata are known or can be
predicted, one approach to steer treatment assignments by someone 
acting fradulently would be to intentionally misclassify
patients' strata.  
Such manipulations are not covered by 
the ignorability conditions.  
This  and other manipulations utlizing knowledge of future treatment codes
are excluded by the usual assumptions for causal inference
in randomized studies (e.g., strongly ignorable treatment assignment), 
which were noted in Section \ref{nonignsect}
, 
so they are not included in the definition of
ignorable/non-ignorable misclassification of strata.  
Analyzing data
using the incorrect strata does not address bias from such manipulation.  
It can be best addressed
by well-designed randomization systems subject to internal and external audits.

\appendix
\section{Appendix:  Derivation of likelihood form in (\ref{uform}) under 
ignorable misclassification}

\label{appA}

The likelihood form in (\ref{uform}) follows from the representation in (\ref{lik3}) provided we
show that 
\begin{equation}
\Prp{ \ytvec \mid \tvec, \svec, \xvec, \svecs,  \muvec}=\Prp{ \ytvec \mid \tvec, \svec, \xvec, \muvec}\ .
\label{dropss}
\end{equation}
The conditional independence of $\ytvec$ and $\svecs$ in equation (\ref{dropss}) is demonstrated here 
using the ignorability assumption (\ref{ignore}), and the independence of all other 
variables with treatment assignment conditional on $\svecs$ for 
permuted block randomizations that depend only on $\svecs$:
$(\ybltvec,\svec,\xvec) \indep \tvec \mid \svecs$.  The probability calculations condition
on the parameters throughout, so they are excluded from the notation in the appendix.  
We first show the potential outcomes
are independent of treatment assignment given $\svec, \xvec$:
\begin{equation}
\ybltvec \indep \tvec \mid \svec, \xvec \ .
\label{interm}
\end{equation}

\noindent Conditioning on $\svecs$,
\begin{eqnarray*}
\Prp{\ybltvec,\tvec \mid \svec,\xvec}&=&
\int \Prp{\ybltvec,\tvec \mid \svec,\xvec,\svecs}d\Prp{\svecs\mid\svec,\xvec}\\
&=&\int\left({\Prp{\ybltvec,\svec,\tvec \mid \xvec,\svecs}
\over
\Prp{\svec\mid\xvec, \svecs}}\right)  d\Prp{\svecs\mid\svec,\xvec}\ .
\end{eqnarray*}
Using the conditional independence of $\tvec$ due to the randomization design, this becomes
\begin{displaymath}
\int\left({\Prp{\ybltvec,\svec \mid \xvec,\svecs} \Prp{\tvec\mid \xvec,\svecs}
\over
\Prp{\svec\mid \xvec, \svecs}}\right)d\Prp{\svecs\mid\svec,\xvec}\\
\ .
\end{displaymath}
Re-expressing the probability on the left side of the numerator yields
\begin{displaymath}
\int\left({\Prp{\ybltvec \mid \svec, \xvec,\svecs}\Prp{\svec \mid \xvec,\svecs} 
\Prp{\tvec\mid \xvec,\svecs}
\over
\Prp{\svec\mid\xvec,\svecs}}\right)d\Prp{\svecs\mid \svec, \xvec}\\
\ .
\end{displaymath}
Using the ignorability assumption followed by the design assumption 
yields the conditional independence in (\ref{interm}):
\begin{eqnarray*}
\Prp{\ybltvec,\tvec \mid \svec,\xvec}
&=&
\int  \Prp{\ybltvec \mid \svec, \xvec}
\Prp{\tvec\mid \xvec,\svecs} d\Prp{\svecs \mid \svec, \xvec} \\
&=&
\Prp{\ybltvec \mid \svec, \xvec}\int  
\Prp{\tvec\mid \svec, \xvec, \svecs} d\Prp{\svecs \mid \svec, \xvec} \\
&=&
\Prp{\ybltvec \mid \svec, \xvec}\Prp{\tvec\mid \svec, \xvec}
\ . 
\end{eqnarray*}

To establish (\ref{dropss}), note that $\ytvec$ is a function of 
$\ybltvec$ given $\tvec$, so it suffices to show that 
$\ybltvec \indep \svecs \mid \tvec, \svec, \xvec$.  
Factoring the joint probability,
\begin{eqnarray*}
\Prp{\ybltvec,\svecs \mid \tvec, \svec,\xvec}
&=&
\Prp{\ybltvec \mid \tvec, \svec, \xvec, \svecs } 
\Prp{\svecs \mid \tvec, \svec, \xvec } 
\\
&=&
{\Prp{\ybltvec, \tvec, \svec \mid  \xvec, \svecs } 
\Prp{\svecs \mid \tvec, \svec, \xvec }
\over
\Prp{\tvec, \svec \mid \xvec, \svecs }
} \ .
\end{eqnarray*}
Applying the design assumption, this becomes
\begin{eqnarray*}
{\Prp{\tvec \mid  \xvec, \svecs }
\Prp{\ybltvec, \svec \mid  \xvec, \svecs } 
\Prp{\svecs \mid \tvec, \svec, \xvec }
\over
\Prp{\tvec\mid \xvec, \svecs }\Prp{\svec \mid \xvec, \svecs }
}&=&\\
\Prp{\ybltvec \mid \svec,  \xvec, \svecs }
\Prp{\svecs \mid \tvec, \svec, \xvec }
 \ .
\end{eqnarray*}
Finally, applying the ignorability assumption and (\ref{interm}) yields

\begin{eqnarray*}
\Prp{\ybltvec,\svecs \mid \tvec, \svec, \xvec}
&=&
\Prp{\ybltvec \mid \svec,  \xvec}
\Prp{\svecs \mid \tvec, \svec, \xvec }\\
&=&
\Prp{\ybltvec \mid \tvec, \svec, \xvec}\Prp{\svecs \mid \tvec, \svec, \xvec}
 \ .
\end{eqnarray*}

\begin{table}[h]
\centering
\begin{tabular}{ccc|ccc|ccc}
\hline  
\multicolumn{2}{c}{Misclassification}&&\multicolumn{3}{c}{Corrected strata}&
\multicolumn{3}{c}{Incorrect strata}\\
\vspace{-2ex}
Low&High&Within&&Lik&\multicolumn{1}{c}{Mean}&&Lik&Mean\\
Strata&Strata&Strata&Bias&Cov&\multicolumn{1}{c}{SE}&Bias&Cov&SE\\
\hline
\multicolumn{2}{c}{Ignorable}&corr&\\
\hline
$\gamma_L=0.02$&$\gamma_H=0.02$&1.0&
0&0.95&
0.307&
0&0.95&
0.31\\
$\gamma_L=0.02$&$\gamma_H=0.02$&0.5&
0&0.95&
0.307&
0&0.95&
0.31\\
$\gamma_L=0.15$&$\gamma_H=0.30$&1.0&
0.001&0.951&
0.308&
0.001&0.95&
0.332\\
$\gamma_L=0.15$&$\gamma_H=0.30$&0.5&
0&0.95&
0.307&
0&0.95&
0.317\\
\hline
\multicolumn{2}{c}{Non-ignorable[1]}&Corr&\\
\hline
$\gamma_L=0.02$&$\gamma_H=0.02$&1.0&
0&0.951&
0.307&
0.001&0.95&
0.3\\
$\gamma_L=0.02$&$\gamma_H=0.02$&0.5&
0.001&
0.951&
0.307&
0&
0.95&
0.304\\
$\gamma_L=0.15$&$\gamma_H=0.30$&1.0&
0&0.985&
0.309&
0&0.95&
0.28\\
$\gamma_L=0.15$&$\gamma_H=0.30$&0.5&
0.001&
0.955&
0.307&
0&
0.956&
0.304\\
\hline
\multicolumn{2}{c}{Non-ignorable[2]}&Corr&\\
\hline
$\gamma_L=0.02$&$\gamma_H=0.02$&1.0&
0.001&0.952&
0.307&
0.001&0.95&
0.318\\
$\gamma_L=0.02$&$\gamma_H=0.02$&0.5&
0&
0.951&
0.307&
0.001&
0.95&
0.315\\
$\gamma_L=0.15$&$\gamma_H=0.30$&1.0&
0&
0.985&
0.309&
0&
0.949&
0.328\\
$\gamma_L=0.15$&$\gamma_H=0.30$&0.5&
-0.002&
0.954&
0.307&
-0.001&
0.949&
0.325\\
\hline
\end{tabular}
\caption
{Summary of estimation from the simulation study.  
`Non-ignorable' refers
to the first [1] and second [2] non-ignorable data generation models. 
The $\gamma_L$ is the misclassification rate of patients in the low strata, and
$\gamma_H$ is the misclassification rate of patients in the high strata.  
`Corr' is correlation
between potential outcomes.  The `Lik Cov' is the coverage of 
the likelihood-based intervals. 
The mean of the $SE$ without misclassification is 
0.307.  
}
\label{simtab}
\end{table}

\begin{table}[h]
\centering
\begin{tabular}{ccc|cccc|cccc}
\hline  
\multicolumn{2}{c}{Misclassification}&&\multicolumn{4}{c}{Corrected strata}
&\multicolumn{4}{c}{Incorrect strata}\\
\vspace{-2ex}
Low&High&&Lik&RB&Lik&
\multicolumn{1}{c}{RB}&\multicolumn{1}{c}{Lik}&
RB&Lik&RB\\
Strata&Strata&&Lev&Lev&Pow&
\multicolumn{1}{c}{Pow}&\multicolumn{1}{c}{Lev}&
Lev&Pow&Pow\\
\hline
\multicolumn{2}{c}{Ignorable}&Corr&\\
\hline
$\gamma_L=0.02$&$\gamma_H=0.02$&1.0&
0.051&0.05&
0.78&0.77&
0.052&0.052&
0.78&0.77\\
$\gamma_L=0.02$&$\gamma_H=0.02$&0.5&
0.047&0.049&
0.78&0.77&
0.047&0.048&
0.77&0.77\\
$\gamma_L=0.15$&$\gamma_H=0.30$&1.0&
0.051&0.052&
0.78&0.78&
0.052&0.054&
0.72&0.71\\
$\gamma_L=0.15$&$\gamma_H=0.30$&0.5&
0.048&0.049&
0.78&0.77&
0.047&0.046&
0.71&0.71\\
\hline
\multicolumn{2}{c}{Non-ignorable[1]}&Corr&\\
\hline
$\gamma_L=0.02$&$\gamma_H=0.02$&1.0&
0.043&0.045&
0.78&0.78&
0.046&0.047&
0.83&0.82\\
$\gamma_L=0.02$&$\gamma_H=0.02$&0.5&
0.047&0.048&
0.79&0.78&
0.047&0.048&
0.81&0.8\\
$\gamma_L=0.15$&$\gamma_H=0.30$&1.0&
0.014&0.049&
0.82&0.91&
0.052&0.053&
0.97&0.97\\
$\gamma_L=0.15$&$\gamma_H=0.30$&0.5&
0.032&0.052&
0.8&0.84&
0.05&0.05&
0.87&0.86\\
\hline
\multicolumn{2}{c}{Non-ignorable[2]}&Corr&\\
\hline
$\gamma_L=0.02$&$\gamma_H=0.02$&1.0&
0.05&0.053&
0.78&0.78&
0.05&0.051&
0.73&0.72\\
$\gamma_L=0.02$&$\gamma_H=0.02$&0.5&
0.053&0.054&
0.79&0.78&
0.053&0.053&
0.75&0.74\\
$\gamma_L=0.15$&$\gamma_H=0.30$&1.0&
0.013&0.048&
0.83&0.92&
0.047&0.048&
0.75&0.74\\
$\gamma_L=0.15$&$\gamma_H=0.30$&0.5&
0.033&0.051&
0.8&0.83&
0.054&0.054&
0.7&0.69\\
\hline
\end{tabular}
\caption
{Summary of testing from the simulation.  The `Lik' and  `RB' refer to likelihood
and randomization based inference, `Lev' and `Pow' refer to the level of the 
nominal $0.05$ test, and its power with $\delta=0.5$.  `Non-ignorable' refers
to the first [1] and second [2] non-ignorable data generation models.  `Corr'
refers to the correlations between the potential outcomes.  
The $\gamma_L$ is the misclassification rate of patients in the low strata, and
$\gamma_H$ is the misclassification rate of patients in the high strata.  
\label{simpowtab}
}
\end{table}

\begin{table}[h]
\centering
\begin{tabular}{c|cccc|cccc}
\hline
\multicolumn{1}{c}{}&\multicolumn{4}{c}{$\corr (y^{(0)},y^{(1)})=1$}&
\multicolumn{4}{c}{$\corr (y^{(0)},y^{(1)})=0.5$}\\
\multicolumn{1}{c}{}&\multicolumn{2}{c}{\svecs=1}&\multicolumn{2}{c}{\svecs=2}
&\multicolumn{2}{c}{\svecs=1}&\multicolumn{2}{c}{\svecs=2}\\
\hline
\multicolumn{9}{c}{High rate ignorable misclassification}\\
\hline  
&Mean&SD&Mean&SD&Mean&SD&Mean&SD\\
\hline
$T=0$&0.35&
1.11&
0.87&
1.05&
0.03&
1.02&
0.91&
1.04\\
$T=1$&0.85&
1.11&
1.37&
1.05&
0.53&
1.02&
1.41&
1.04\\
\hline
\multicolumn{9}{c}{High rate non-ignorable [1] misclassification}\\
\hline
&Mean&SD&Mean&SD&Mean&SD&Mean&SD\\
\hline  
$T=0$&-0.23&
0.72&
1.49&
0.68&
-0.27&
0.81&
1.07&
0.96\\
$T=1$&0.27&
0.72&
1.99&
0.68&
0.31&
0.96&
1.53&
0.94\\
\hline
\multicolumn{9}{c}{High rate non-ignorable [2] misclassification}\\
\hline
&Mean&SD&Mean&SD&Mean&SD&Mean&SD\\
\hline  
$T=0$&0.94&
1.14&
0.24&
0.96&
0.37&
0.9&
0.73&
1.2\\
$T=1$&1.44&
1.14&
0.74&
0.96&
0.78&
1.14&
1.27&
1.07\\
\hline
\end{tabular}
\caption{
Summary of population distributions for the misclassified strata.  
The control means in the correct lower and upper strata are $0$ and $1$.  
The corresponding treated means are $0.5$ and $1.5$.  The SD is $1$ in each
treatment by correct strata group.  The proportion of
patients in the correct lower strata is $0.4$. The proportion of patients in
the misclassified lower strata given in row-wise order are
$0.52$, $0.35$, 
$0.52$, $0.35$
$0.52$, and 
$0.35$. Non-ignorable [1] and [2] refer
to the first and second non-ignorable data generation models.} 
\label{poptab}
\end{table}

\section{Appendix:  Extending \cite{bugni18} to subgroup analyses}
\label{appb}

This appendix provides additional details showing that when a randomization
plan satisfies the conditions 2(a) and 2(b) of \cite{bugni18}, it also 
satisfies these conditions on a pre-specified subgroup defined by
baseline patient characteristics.  This implies subgroup analyses that 
do not adjust for 
strata when applied to designs with randomized blocks within 
strata are conservative.  
Of special interest are subgroups corresponding to corrected strata in
settings where the randomization was based on initial strata that included 
misclassified values.

As in \cite{bugni18}, the result is derived for the setting with a control 
group and a single treatment, which is denoted by `1'.  The strata, other
covariates, and potential outcomes are also assumed to be independent
and identically distributed as in the reference, which was not required
for the general formulation in Section \ref{ignsect}.  The notation in 
Section \ref{ignsect} is extended to included strata and subgroup counts.  
The number of patients in strata $j$ is denoted by $N_j$, and the number
of treated patients in strata $j$ is denoted by $N_{j1}$.  Using the notation
of \cite{bugni18}, the targeted proportion of treated patients in each
strata is $\pi$.  The proportion of treated
patients actually randomized in each strata is denoted by $\pihat_j=N_{j1}/N_j$.  
Assumption 2(b), which is satisified by stratifed permuted block designs,
assures that the $\var(\pihat_j)=\tau_j/N_j$, with $\tau_j\leq 
\pi(1-\pi)$.  A new variable is defined as, $R_i=1$, when a patient is 
in the pre-specified
subgroup, and $0$ otherwise.  The number of patients in the subgroup in
strata $j$ is denoted by $N_{Rj}$, the corresponding number of treated
patients is denoted by $N_{Rj1}$, and the proportion of such patients is
denoted by $\pihat_{Rj}=N_{Rj1}/N_{Rj}$.

A key feature of the stratified permuted block design and the other designs
considered by \cite{bugni18} is that conditional on $S_i$, all baseline
covariates including the subgroup indicator $R_i$ are independent of the
treatment assignments, $T_i$.  Amongst the patients with $S_i=j$, the 
number of patients in the strata are known, and the number of patients
in the subgroup, $N_{Rj}$, is independent of the number of treated patients
in the strata, $N_{j1}$.  The remaining calculations condition on
the full set of $S_i$ through the $N_j$.  Further conditioning on $N_{j1}$ and $N_{Rj}$,  
the independence of $R_i$ and $T_i$, and the independent identically 
distributed assumption for the covariates imply all subsets of $N_{Rj}$
patients are equally likely to be the selected subgroup, 
i.e., the $N_{Rj}$ patients
are a random sample without replacement from all patients in the strata.  
Note that the representation does not require independence between the
$T_i$.  The expectation and variance of the sample proportion is 
\citep{sarndal92}:
\begin{equation}
\E \left(\pihat_{Rj}\mid N_{Rj},N_{j1},N_j\right)=\pihat_j \ ,
\label{cmean}
\end{equation}
and 
\begin{equation}
\var \left(\pihat_{Rj}\mid N_{Rj},N_{j1},N_j\right)=
\frac{\pihat_j(1-\pihat_j)}{N_{Rj}}
\left(1-N_{Rj}/N_j\right)\ .
\label{cvar}
\end{equation}
Averaging over $N_{j1}$, 
$\E\left(\pihat_{Rj}\mid N_{Rj}, N_j\right)=\pi$.  
The variance
is computed from the sum of $\var(\pihat_j\mid N_{Rj},N_j)=\tau_j/N_j$,  
and the expectation of (\ref{cvar}).  Because the $\pihat_j$ are bounded
and converge in probability to $\pi$, the expectation of (\ref{cvar}) is
asymptotically \citep{billingsley}:
\begin{equation}
\var \left(\pihat_{Rj}\mid N_{Rj}, N_j\right)=
\frac{\pi(1-\pi)}{N_{Rj}}
\left(1-N_{Rj}/N_j\right)\ ,
\label{cvarE}
\end{equation}
so
\begin{eqnarray}
\var \left(\pihat_{Rj}\mid N_{Rj}, N_j\right)&=&
\frac{\tau_j}{N_j} + 
\frac{\pi(1-\pi)}{N_{Rj}}
\left(1-N_{Rj}/N_j\right) \nonumber\\
&=& 
\frac{\tau_j}{N_j}+\frac{\pi(1-\pi)}{N_{Rj}} 
- \frac{\pi(1-\pi)}{N_{j}}\nonumber\\
&=&
\frac{\pi(1-\pi)}{N_{Rj}}+ \left(\tau_j-\pi(1-\pi)\right)/N_j 
\ . \label{ineqB}
\end{eqnarray}
The rightmost term in (\ref{ineqB}) is $\leq 0$, showing that
the subgroup-specific strata treatment rates also satisfy the
criteria for conservative inference.

{
\setlength{\baselineskip}{12pt}


}

\end{document}